\documentclass[12pt]{article}
\usepackage{amsmath}
\usepackage{graphicx}
\usepackage{amssymb}
\usepackage{float}
\usepackage[bottom=1 in,top=1.3 in]{geometry}
\begin{document}
\title{\textbf{Study of effective coupling between charge degrees of freedom in low dimensional hole-doped quantum antiferromagnets}}
\author{\textbf{Suraka Bhattacharjee$^1$} and \textbf{Ranjan Chaudhury$^{1,2}$} \\ 
$^1$Department of Condensed Matter Physics and Material Sciences\\
S.N. Bose National Centre for Basic Sciences \\ Saltlake, Sector-III, Block-JD, Kolkata-700106, India\\
$^2$Department of Physics\\
Ramakrishna Mission Vivekananda Educational and Research Institute \\
Belur-711202, India \\
Email- $^1$surakabhatta@bose.res.in, $^{1}$ranjan@bose.res.in,\\ $^{2}$ranjan.chowdhury@rkmvu.ac.in}
\maketitle

\section*{Abstract}

Expressions for generalized charge stiffness constant at zero temperature are derived corresponding to low dimensional hole doped quantum antiferromagnets, describable by the t-J-like models, with a view to understanding fermionic pairing possibilities and charge couplings in the itinerant antiferromagnetic systems. A detailed comparison between spin and charge correlations and couplings are presented in both strong and weak coupling limits. The result highlights that the charge and spin couplings show very similar behaviour in the over-doped region in both the dimensions, whereas they show a completely different trend in the lower doping regimes. A qualitative equivalence of generalized charge stiffness constant with the effective Drude weight and Coulomb interaction is established based on the comparison with other theoretical and experimental results. The fall in charge stiffness with increase in doping then implies reduction in the magnitude of effective Coulomb repulsion between the mobile carriers. This leads to an enhanced possibility of fermionic pairing with increase in doping in the possible presence of some other attraction producing mechanism from a source outside the t-J-like models. Moreover, under certain conditions in the weakly correlated phase, the t-J-like models themselves are able to produce attractive interaction for pairing.
%\keywords{generalized charge stiffness constant \and charge coupling \and Drude weight \and effective Coulomb interaction \and t-J-like models \and fermionic pairing}

\section{Introduction}
\label{intro}
Most of the layered cuprate superconductors are known to exhibit many characterististic phases, supported by consistent experimental evidences \cite{1,2,3}. The spin dynamics plays an important role in studying the magnetic behaviour of the phases, bearing the signatures of strong and weak correlations in the different doping regions. The phases include the long range ordered antiferromagnetic phase in low doping regime, anomalous non-Fermi liquid-like conducting phase and normal Fermi liquid-like conducting phase at higher doping regions. Interestingly, the optimally doped region shows high temperature superconductivity below the corresponding critical temperature \cite{1,2,3}. However, the subsequent discussions about this unconventional superconductivity in cuprates are necessarily accompanied by the possibility of pair formation in these systems. The interaction between the charge degrees of freedom, in effect to the Coulomb potential, are important in determining the pairing possibility in the strongly correlated doped phases \cite{4}. Study of correlations between the spin and charge degrees of freedom in the itinerant phases of doped cuprates involves the Cu and the O bands \cite{5,6}. Later, the two band Hamiltonian was reduced to the well known single band t-J model in the low energy limit \cite{7,8,9}.\\
\hspace*{0.3cm} The magnetic interaction in 2D systems was studied using  many theoretical approaches including  Mori's projection technique based on two-time thermodynamic Green's function and Variational Monte Carlo simulations \cite{10,11,12,13,14}. On the other hand, the 1D t-J model is exactly solvable using Bethe Ansatz at specific values of J/t \cite{15,16}. Density Matrix Renormalization Group (DMRG) and Transfer Matrix Renormalization Group (TMRG) techniques have been used very successfully in 1D to find the spin correlations away from the super-symmetric points \cite{17,18}. In 2D too, some attempts have been taken using DMRG to find the spin and charge density orders in the doped Hubbard model \cite{19}. In our recent papers, we have developed a non-perturbative  quantum mechanical approach to determine the spin correlations in both 2D and 1D doped antiferromagnets, on the basis of generalized spin stiffness constant corresponding to the t-J model \cite{20,21}. Our results in 1D lead to a very interesting consequence regarding the formation of a new type of spin-spin coupling as doping increases, which is totally distinct from the original antiferromagnetic coupling seen in the insulating and under-doped phases \cite{21}. Our novel prediction was further supported by other experimental and theoretical results \cite{21}.  \\
\hspace*{0.3cm} Beside the spin correlations, the attempts to determine the charge correlations include the determination of the inverse dielectric function, involving the standard many body formalism in a Fermi liquid \cite{22}. The total free energy used in the calculation comprises of the Hartree-like term and the exchange correlation contributions. It was found that the Coulomb interaction thus calculated from the inverse of dielectric function, can even change sign and turn attractive if the spin susceptibility is larger than a threshold value \cite{22}. This can trigger the possibility of pairing in some of the doped antiferromagnetic systems. However, the above technique could not determine the charge coupling in strongly correlated phases of the systems, where the double fermionic occupancy on each site is disallowed. \\
\hspace*{0.3cm} The other approaches include the finding of the local charge stiffness tensor (D$_{\alpha\beta}$) as the response of the system to any change in boundary condition \cite{23}. The component D$_{\alpha\alpha}$ was used to find the optical mass and was shown to be directly proprtional to the Drude weight \cite{23}. But the magnitudes of charge stiffness constants, calculated by applying the Lanczos algorithm, were determined only at discrete values of hole concentrations \cite{23,24}. The Drude weight calculated by exact diagonalization technique in Hubbard cluster shows an increase in the lower doping regime, where the interacting holes are considered as the major carriers \cite{25}. Furthermore, in the over-doped regime, the weakly interacting electrons take the role of the major carriers and the Drude weight falls in magnitude \cite{26}.  Moreover, the dynamical conductivity derived based on the memory function technique in terms of the Hubbard operators, was found to be proportional to doping concentration \cite{9}. In contrast to the 2D case, both Hubbard and t-J models are exactly solvable in 1D, involving the Bethe ansatz \cite{27,28,29,30}. The transport properties for the 1D Hubbard model has been studied using the Bethe Ansatz solution combined with the global symmetry and the operator algebra for the Hubbard operators \cite{31}. The charge stiffness constant calculated at finite temperature (T$>$0) corresponds to the response to a static field characterizing the weightage of the Drude peak \cite{31}. However, these calculations were carried out only on the exactly half-filled Hubbard model i.e., zero doping limit. In order to have a more clearer, definite and detailed understanding of the doping dependences of the charge stiffness, we embark upon an analytical approach.   \\
\hspace*{0.3cm} In this paper, our main aim would be to determine the interaction and coupling between the charge degrees of freedom and to put forward a comparative study between the charge and the spin couplings. Similar to the case of spin degrees of freedom, here the doping dependence of charge-charge coupling is studied in terms of the evolution of generalized charge stiffness constant with doping concentration at T=0. In the strongly correlated under-doped regime, we have involved the nearest neighbour t-J model preventing the double occupancies. However, in the weakly correlated over-doped regime, we have used the t$_1$-t$_2$-t$_3$-J model with the Gutwiller variational parameter $\alpha$ very small or zero, which allows double occupancies in the system. The results of charge stiffness in the lower doping regions are compared with other theoretical and experimental results on layered cuprate systems \cite{26,32}. Based on the comparisons, we have shown a qualitative equivalence between Drude weight and our derived charge stiffness constant. The connection between charge stiffness and effective Coulomb interaction in the doped regimes is also established within the framework of random phase approximation (RPA). Finally, we have explored the consequences and various possibilities arising from our systematic studies as stated above.

\section{Results}
\subsection{Calculational formalism and numerical results for charge stiffness}
\subsubsection{Strongly correlated and with nearest neighbour hopping}
The nearest neighbour t-J model Hamiltonian for strongly correlated electronic systems is \cite{33,34}:
\begin{align}
                   H_{t-J}=H_t+H_J
\end{align}
where H$_t$ and H$_J$ represents the hopping and exchange interactions involving nearest neighbour sites, respectively with restrictions on double occupancy at each site. The expression for the kinetic energy Hamiltonian is given as \cite{33,34}:
\begin{align}                                  
 H_t=-\sum_{<i,j>,\sigma}t_{ij}X^{\sigma0}_iX^{0\sigma}_j
\end{align}
Here t$_{ij}$ represents the hopping amplitude from j$^{th}$ to i$^{th}$ site and for nearest neighbour t$_{ij}$=t and the X's are the Hubbard operators. \\ 
Again for the exchange energy part is represented as\cite{33,34}:
\begin{align}
H_J=\sum_{<ij>}J_{ij}(\overrightarrow{S_i}.\overrightarrow{S_j}-\frac{1}{4}n_i n_j)
\end{align}
where S$_i$ and S$_j$ now represent the localized spin operators corresponding to the i$^{th}$ and j$^{th}$ sites respectively; J$_{ij}$ is the exchange constant involving the i$^{th}$ and the j$^{th}$ site and for nearest neighbour pair $\langle$ij$\rangle$, J$_{ij}$=J; n$_i$ and n$_j$ are the occupation number operators for the i$^{th}$ and j$^{th}$ site respectively.\\
As was done earlier for generalized spin stiffness constant ($\tilde{D_s}$), a similar kind of equation also holds for the generalized charge stiffness ($\tilde{D_c}$)
 \begin{align}
 \tilde{D_c}=\tilde{D_c^t}+\tilde{D_c^J}
\end{align}   
where $\tilde{D_c^t}$ and $\tilde{D_c^J}$ are the contributions to charge stiffness constant from kinetic energy and exchange energy respectively and are given by \cite{35,36}:
     \begin{align}
    \tilde{D_c^t}=\lim_{\phi\rightarrow0}(\frac{1}{2})\frac{\delta^2T}{\delta\phi^2}
\end{align} and
\begin{align}
    \tilde{D_c^J}=\lim_{\phi\rightarrow0}(\frac{1}{2})\frac{\delta^2E_J}{\delta\phi^2}
\end{align}
where `T' and `E$_J$' are the kinetic energy expectation value and exchange energy expectation value of the t-J Hamiltonian. $\phi$ is the `electric twist' corresponding to the Peierl's phase $\phi_\sigma$ arising from the presence of the vector potential A($\overrightarrow{r}$) as used in the definition of generalized stiffness constants \cite{35,36}, where the quantity $\phi_\sigma$ has the following property for the charge response (spin symmetric case):
 \begin{align}
 \phi_\downarrow=\phi_\uparrow=\phi
\end{align} 
[This is unlike the spin asymmetric case, where we had used $\phi_\downarrow=-\phi_\uparrow=\phi$  \cite{20,21,36}]\\
\hspace*{0.3cm} We have evaluated the expectation values in the Gutzwiller state.
\begin{align}
\vert\psi_G\rangle=\prod_l(1-\alpha\widehat{n}_{l\uparrow}
 \widehat{n}_{l\downarrow})\vert{FS}\rangle
\end{align}
with $\alpha$ as the variational parameter deciding the amplitude for no-double occupancy of any site and $\vert$FS$\rangle$ is the Fermi sea ground state \cite{20,21,36}. At first we take $\alpha$=1 for completely projecting out the doubly occupied sites.
\begin{align}
\vert\psi_G\rangle_{NDOC}=\prod_l(1-\widehat{n}_{l\uparrow}
 \widehat{n}_{l\downarrow})\prod_{k\sigma}^{k_F}\sum_{ij}
 C_{i\sigma}^{\dagger} C_{j-\sigma}^{\dagger}
 e^{i(\overrightarrow{r_i}-\overrightarrow{r_j}).\overrightarrow{k}}\vert{vac}\rangle
\end{align}
where $\vert{vac}\rangle$, i, j and l have the usual meaning \cite{36}. \\
The exchange energy for the spin symmetric case (see eq.(6)) can be written as:
\begin{align}
E_J=(\frac{zt_{eff}^2}{V_{eff}})
\frac{_{NDOC}\langle\psi_G\vert{H_J'}\vert\psi_G\rangle_{NDOC}}{_{NDOC}\langle\psi_G\vert\psi_G\rangle_{NDOC}}
\end{align}
where `z' is the co-ordination number i.e., z=4 for 2-D and 2 for 1-D and  
\begin{align}
H_J'=\overrightarrow{S_i}.\overrightarrow{S_j}-\frac{1}{4}n_in_j
\end{align}with
$_{NDOC}\langle\psi_G\vert\psi_G\rangle_{NDOC}$ being the normalization factor for the Gutzwiller state $\vert\psi_G\rangle_{NDOC}$ \cite{21}. \\ \\
\hspace*{0.3cm} Since E$_J$ is $\phi$ independent [see eq.(10)],
\begin{align}
  \tilde{D^J_c}=0
\end{align}
\begin{center}
\hspace*{0.3cm} Thus $\tilde{D_c}=\tilde{D_c^t}$ always.\\ \end{center}
Hence the exchange energy contribution to charge stiffness vanishes in the entire doping region. This may be completely physical because the interchange of spins has no effect on the carriers in terms of their charge responses. \\ 
\hspace*{0.3cm} The total charge stiffness is given by the kinetic energy contribution to charge stiffness ($\tilde{D_c^t}$). The kinetic energy is derived as: \\
In 2D,
\begin{align}
T(\phi\neq0)=(-t)[\prod_{k_x,\sigma}^{k_F}\sum_{\sigma}4cos(k_xa)(1-\delta)^2 cos(\phi_\sigma)-N_l\prod_{k_x,\sigma}^{k_F}\sum_\sigma 4cos(k_xa) cos(\phi_\sigma)/N^2]
\end{align}
Now taking the second order derivative, one can get:
\begin{align}
  \tilde{D_c}=(t)[\prod_{k_x,\sigma}^{k_F}4cos(k_xa)(1-\delta)^2-N_l\prod_{k_x,\sigma}^{k_F}4cos(k_xa)/N^2]
\end{align} 
(while the vector potential is applied in x-direction
)\\
Similarly, for 1D,
\begin{align}
  \tilde{D_c}=(t)[\prod_{k,\sigma}^{k_F}4cos(ka)(1-\delta)^2-N_l\prod_{k,\sigma}^{k_F}4cos(ka)/N^2]
\end{align}
where N$_l$=N(1-$\delta$), N is the total number of sites and `$\delta$' is the doping concentration and the Fermi momentum k$_F$ in 2-D has the form in the quasi-continuum approximation \cite{20,21,36}:
 \begin{align}
       k_F=\frac{\sqrt{2\pi(1-\delta)}}{a}
\end{align}
and in 1-D:
\begin{align}
       k_F=(\pi/2a)(1-\delta)
\end{align}
Here it can be noted that the form of $\tilde{D^t_c}$ is similar to that of $\tilde{D^t_s}$ in both one and two dimensions \cite{20,21,36}. Hence following the same arguments described in our two previous papers \cite{20,21},  $\tilde{D_c}$ vanishes if at least one value of k$_x$ in 2D (k in 1D) satisfies:\\
For 2D,
\begin{align}
k_xa=\pi/2
\end{align}
and for 1D
\begin{align}
ka=\pi/2 
\end{align}
This condition can be satisfied only when k$_F$a=$\pi$/2. Using the expressions for k$_F$ (see eqs. (16,17)), one can get the vanishing conditions are $\delta\rightarrow1$ and $\delta\leqslant$0.61 for 2D model and at $\delta\rightarrow1$ and $\delta\rightarrow0$ for 1D \cite{20,21,36}. For the vector potential applied in the x-direction, we get the value of $\delta$=$\delta_c\approx$0.61, below which the charge stiffness remains zero in 2D.  \\
\hspace*{0.3cm}  The total charge stiffness constants derived for the strongly correlated $\alpha$=1 case in 2D and 1D are plotted against $\delta$ (see Figs.(1,2)) \cite{20,21}. In the plots, the total charge stiffness has been scaled down by the number of pairs of mobile holes in the system, to extract an equivalent stiffness  corresponding to a pair of mobile charge carriers: 
%\vspace*{-1cm} 
  \begin{align}
  D_c=\tilde{D_c}/^{N_{l}}C_2
\end{align} 
%\vspace*{0.5cm} 
\begin{figure}[H]
%\begin{subfigures}
\centering
\minipage{0.48\textwidth}
\fbox{{\includegraphics[width=\linewidth]{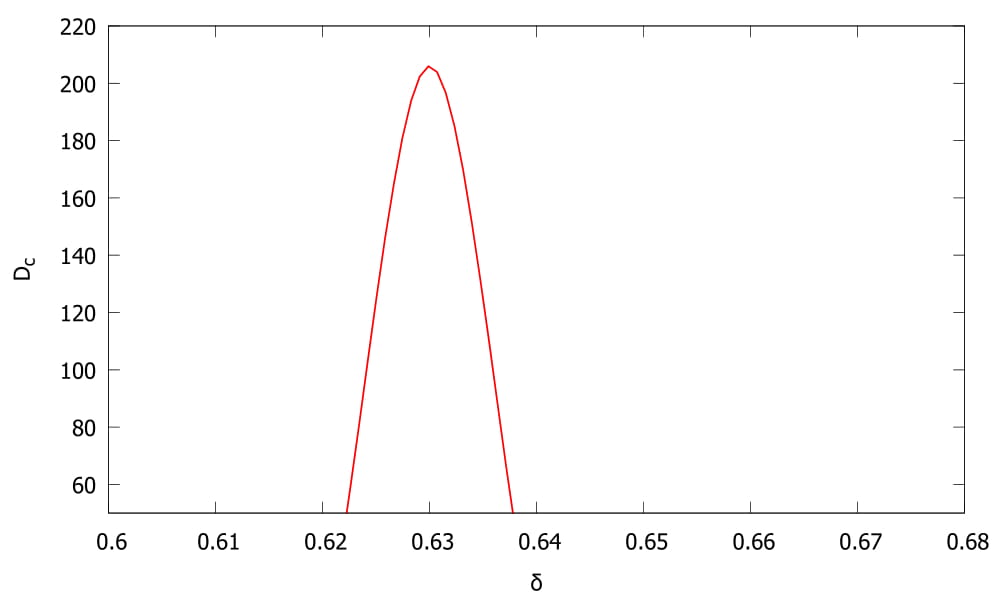}}}
\endminipage\hfill
\minipage{0.48\textwidth}
\fbox{{\includegraphics[width=\linewidth]{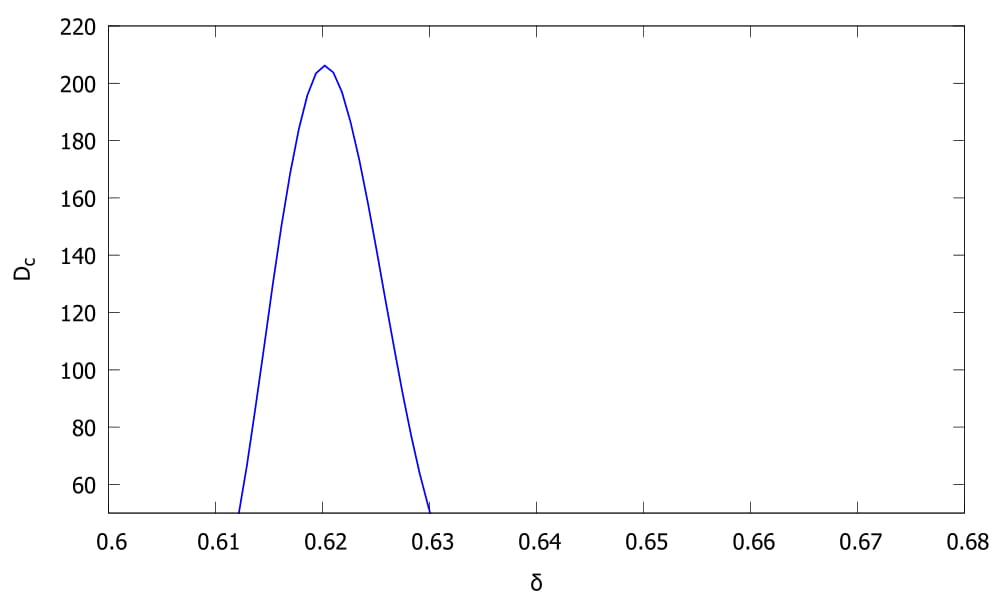}}}
\endminipage\hfill
%\end{subfigures}
\caption{ D$_c$ vs. $\delta$ in 2D: {(a)} lattice size=700x700; {(b)} lattice size=800x800}
 \end{figure}
 
\begin{figure}[H]
%\begin{subfigures}
\minipage{0.48\textwidth}
\fbox{{\includegraphics[width=\linewidth]{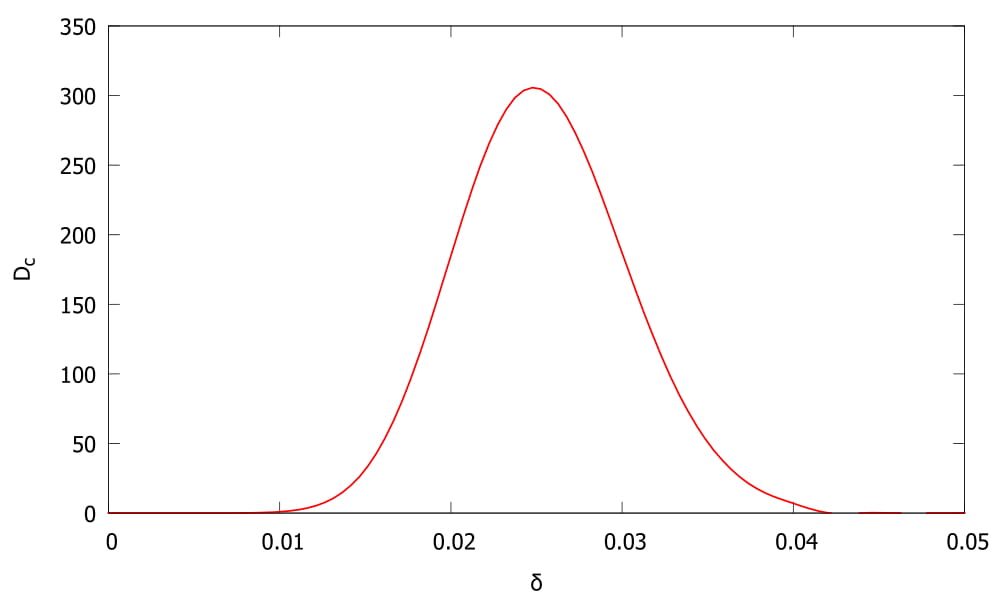}}}
\endminipage\hfill
\minipage{0.48\textwidth}
\fbox{{\includegraphics[width=\linewidth]{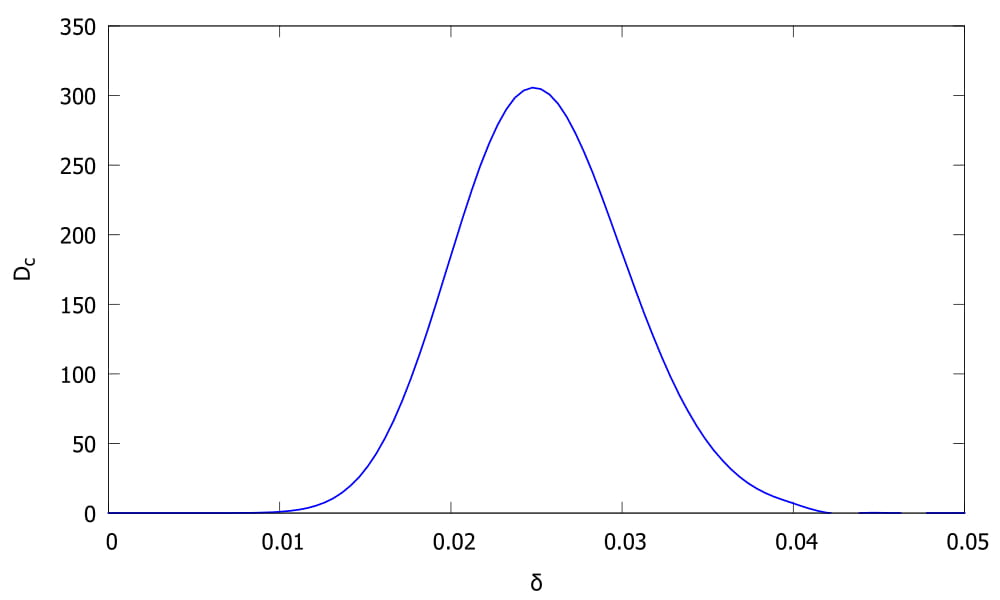}}}
\endminipage\hfill
%\end{subfigures}
\begin{center}
\caption{ D$_c$ vs. $\delta$ in 1D: {(a)} lattice length=1800; {(b)} lattice length=1900}
\end{center}
\end{figure} 

In 2D, the scaled charged stiffness constant vanishes  upto the critical doping concentration $\delta_c$, followed by a sharp rise in D$_c$. The D$_c$ again falls drastically with further increase in doping concentration, giving rise to the appearance of a very sharp cusp-like peak in the over-doped region as shown in Figs.(1a,b). For the 1D model, D$_c$ shows a maximum in the low doping region, and zero elsewhere (see Fig.(2a,b)).\\
 Nevertheless, the calculation in the over-doped regime is not justified only with the nearest neighbour t-J model. The inclusion of higher neighbour hopping terms are necessary for correctly predicting the behaviour of the higher doping regions. 
\subsubsection{Weakly correlated with higher neighbour hoppings}
In the previous sub-section, we have derived the charge stiffness in the strongly correlated regime,
considering only the nearest neighbour interaction. Now, in this sub-section we will consider the over-doped regime with very small $\alpha$ and $\alpha$=0 i.e, allowing double occupancies in the system. Moreover, as we have already stated that in the over-doped regime, the higher neighbour hoppings are also significant, so we have incorporated two higher neighbour terms in the t-J model. \\
The t$_1$-t$_2$-t$_3$-J model is given as \cite{37}:
\begin{align}
H=-t_1\sum_{<i,j>,\sigma}C^{\dagger}_{i\sigma}C_{j\sigma} -t_2\sum_{<<i,j>>,\sigma}C^{\dagger}_{i\sigma}C_{j\sigma}-t_3\sum_{<<<i,j>>>,\sigma}C^{\dagger}_{i\sigma}C_{j\sigma}+J\sum_{<i,j>,\sigma}S_i.S_j
\end{align}
where t$_1$, t$_2$ and t$_3$ represent the first, second and third neighbour hopping amplitudes respectively.\\
With the vector potential applied along the x-direction as before, we get, in 2D,
 \begin{align}
\notag \tilde{D}_c=[\prod_{k_x,\sigma}^{k_F}4\lbrace (t_1)cos(k_xa)+(t_2)cos(2k_xa)+(t_3)cos(3k_xa)\rbrace (1-\delta)^2-  \\
\alpha N_l\prod_{k_x,\sigma}^{k_F}4\lbrace (t_1)cos(k_xa)+(t_2)cos(2k_xa)+(t_3)cos(3k_xa)\rbrace/N^2]
\end{align}
and in 1D,
\begin{align}
\notag \tilde{D}_c=[\prod_{k,\sigma}^{k_F}4\lbrace (t_1)cos(ka)+(t_2)cos(2ka)+(t_3)cos(3ka)\rbrace (1-\delta)^2-  \\
\alpha N_l\prod_{k,\sigma}^{k_F}4\lbrace (t_1)cos(ka)+(t_2)cos(2ka)+(t_3)cos(3ka)\rbrace/N^2]
\end{align}
Now, we consider the limiting case with $\alpha$=0 i.e, the double occupancy is totally allowed on the sites and then the Gutzwiller state reduces to that of an ideal Fermi system: 
\begin{align}
\vert FS\rangle=\prod_{k\sigma}^{k_F}\sum_{ij}
 C_{i\sigma}^{\dagger} C_{j-\sigma}^{\dagger}
 e^{i(\overrightarrow{r_i}-\overrightarrow{r_j}).\overrightarrow{k}}\vert{vac}\rangle
\end{align}
Calculating the kinetic energy in this case ($\alpha$=0) we get for 2D,
\begin{align}
 \tilde{D}_c=\prod_{k_x,\sigma}^{k_F}4\lbrace (t_1)cos(k_xa)+(t_2)cos(2k_xa)+(t_3)cos(3k_xa)\rbrace (1-\delta)^2 
\end{align}
and for 1D,
\begin{align}
\tilde{D}_c=\prod_{k,\sigma}^{k_F}4\lbrace (t_1)cos(ka)+(t_2)cos(2ka)+(t_3)cos(3ka)\rbrace (1-\delta)^2
\end{align}

From eqs.(22-26), one can see that the vanishing conditions for $\tilde{D}_c$ corresponding to very small $\alpha$ and $\alpha$=0 in 2D are $\delta\rightarrow$1 and $\delta\leqslant\delta_c$, where $\delta_c$ depends on the relative magnitudes of t$_1$, t$_2$ and t$_3$. For t$_2$=t$_3$=0, the value of $\delta_c$ goes to 0.61, which is exactly the same as the corresponding value of $\delta_c$ obtained for the nearest neighbour t-J model. For 1D t$_1$-t$_2$-t$_3$-J model, the point, where the stiffness exhibits a jump ($\delta_c$) appears in the optimal doping region which is much lower than that was obtained from the nearest neighbour t-J model. The charge stiffness again falls with further increase in doping concentration, due to the the presence of large number of vacancies in the system. The recent experimental observations from some of the doped correlated systems seem to have a link with this result of ours \cite{37}. \\
\hspace*{0.3cm} The plots of D$_c$ for weakly correlated t$_1$-t$_2$-t$_3$-J model in two dimension, are presented in Fig.(3). The corresponding plots for 1D are given in Fig.(4). The values of t$_2$/t$_1$ and t$_3$/t$_1$ were determined by fitting the tight binding Fermi surfaces to the experimental results on La$_{2-x}$Sr$_x$CuO$_4$ and Bi2212 \cite{39,40}. The second neighbour hopping amplitude was found to be of opposite sign with respect to the first neighbour hopping. Here, we have done the calculations for a range of feasible values of t$_2$ and t$_3$ and presented a result for a few sets of t$_2$/t$_1$ and t$_3$/t$_1$. 
%\vspace*{3cm}
\begin{figure}[H]
\centering
\fbox{\includegraphics[scale=0.6]{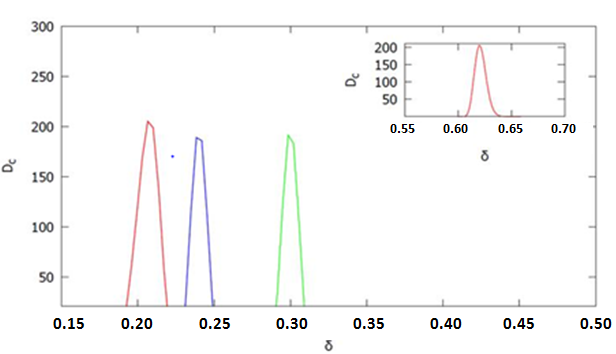}} \caption{D$_c$ vs. $\delta$ for 2D t$_1$-t$_2$-t$_3$-J model, with $\alpha$=0; (a)peak at $\delta\sim$0.29 (t$_2$=-0.53t$_1$,t$_3$=0.24t$_1$) [green line]; (b)peak at $\delta\sim$0.23 (t$_2$=-0.52t$_1$,t$_3$=0.45t$_1$) [blue line]; (c)peak at $\delta\sim$0.19 (t$_2$=-0.6t$_1$,t$_3$=0.56t$_1$) [red line] [in the inset is shown D$_c$ vs. $\delta$ for t$_2$=t$_3$=0; the peak is seen at $\delta\sim$0.61]}
 \end{figure}

\vspace{-1.2cm}
\begin{figure}[H]
\centering
\fbox{\includegraphics[scale=0.27]{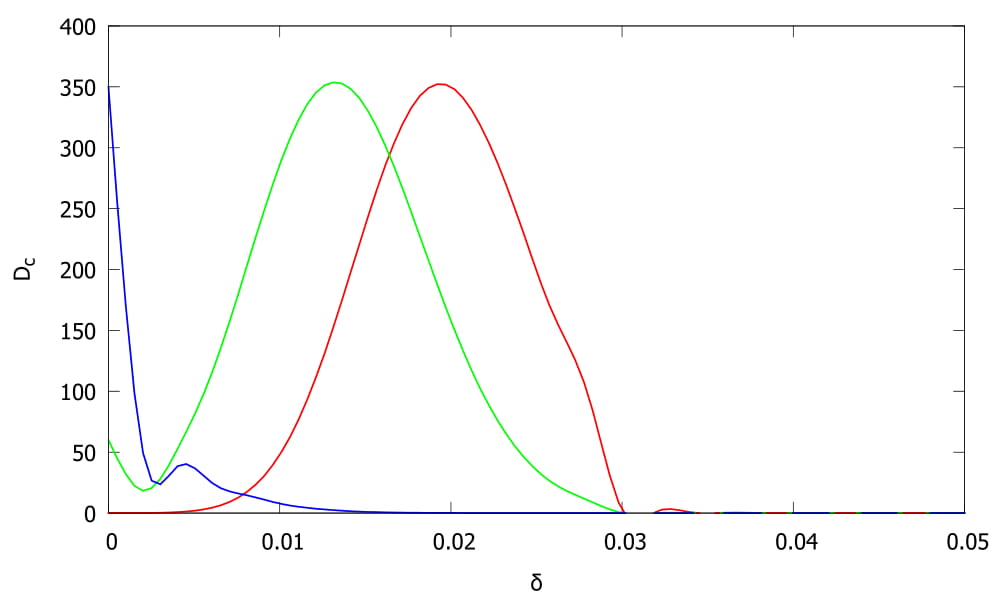}} \caption{D$_c$ vs. $\delta$ for 1D t$_1$-t$_2$-t$_3$-J model, with $\alpha$=0; (a)peak at $\delta\sim$0.02 (t$_2$=-0.01t$_1$,t$_3$=0.005t$_1$) [red line]; (b)peak at $\delta\sim$0.013 (t$_2$=-0.02t$_1$,t$_3$=0.01t$_1$) [green line]; (c)peak at $\delta\rightarrow$0 limit (t$_2$=-0.04t$_1$,t$_3$=0.02t$_1$) [blue line]}
\end{figure}

\vspace*{-0.5cm}

The Fig.(3) shows that the maximum in D$_c$ shifts to the optimal doping region for range of values of t$_2$/t$_1$ and t$_3$/t$_1$. Again, the peak gradually shifts to further lower doping concentration for relatively higher magnitudes of second and third neighbour hopping amplitudes ($\mid t_2\mid$ and $\mid t_3\mid$)(see Fig.(3)). Here one might notice that the magnitude of the scaled charge stiffness is greatly reduced with increase in doping, however, in the previous section we have shown from the analytical calculations, that it quantitatively goes to zero only at 100$\%$ doping concentration, denoting the absence of any carrier in the system.\\
\hspace*{0.3cm}Similarly in 1D too, the peak in D$_c$ shifts to very low doping regime as $\mid t_2\mid$ and $\mid t_3\mid$ are enhanced and the position of the peak reaches $\delta\rightarrow$0 limit at t$_2\approx$-0.04t$_1$ and t$_3\approx$0.02t$_1$ (see (Fig.(4)).

\subsection{Comparison with other theoretical and experimental results}
The imaginary conductivity for the electric field applied in the x-direction is generally derived using the Linear Response Theory as \cite{41,42}:
\begin{equation}
\sigma_{xx}^{\prime \prime}(\omega)=\frac{2e^2}{L^{d}\hbar^{2}\omega}[\frac{1}{2d}\langle -T\rangle-\mathcal{P}\sum_{\nu\neq0}\frac{\mid\langle 0 \mid j_x \mid \nu \rangle \mid^2(E_\nu -E_0)}{(E_\nu -E_0)^2-\hbar^2 \omega^2}]
\end{equation}
where `L' is the lattice length of the `d'-dimensional lattice. $\langle T \rangle$ is the kinetic energy expectation value of the operator T$_x$=-2t$\sum$cos k$_x$C$_k^\dagger$C$_k$ and j$_x$=2t$\sum$sin k$_x$C$_k^\dagger$C$_k$ \cite{41}. `E$_\nu$' and `E$_0$' are the energy eigen values of the $\nu^{th}$ state and the ground state respectively \cite{41}.
In the very low frequency limit, the imaginary conductivity is related to the charge stiffness (D$_c$) by \cite{41,42}:
\begin{equation}
lim_{\omega\rightarrow0}\omega\sigma_{xx}^{\prime \prime}(\omega)=(2e^2/\hbar^2)D_c
\end{equation}
Using the Kramer's Kronig transformation, the real conductivity in the low frequency limit is derived as \cite{41,42}:
\begin{equation}
\sigma_{xx}^{\prime}(\omega)=\frac{2\pi e^2}{\hbar}[D\delta(\hbar\omega)+\frac{1}{L^d}\sum_{\nu\neq0}\mid\langle 0 \mid j_x \mid \nu \rangle \mid^2 \delta((E_\nu -E_0)^2-\hbar^2 \omega^2)]
\end{equation}
`D' is the Drude weight implying the free acceleration of the electrons or dc conductivity. In the low $\omega$ limit, the Drude weight corresponds to the charge stiffness constant (D$_c$) (see eqs.(27)-(28)) \cite{41}.\\
The Drude weight calculated using exact diagonalization for Hubbard model on 4x4 site cluster is shown in Fig.(5) \cite{26}.
\begin{figure}[H]
\begin{center}
\fbox{\includegraphics[scale=0.65]{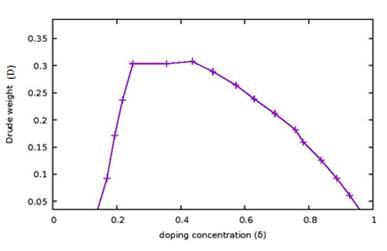}} \caption{Drude weight vs. doping on 4x4 cluster for U/t=4 using exact diagonalization technique [reproduced from ref.(26)}
\end{center}
\end{figure}

From Fig.(5), it can be seen that in the very low doping region, the Drude weight is zero as a result of insulating behaviour of the antiferromagnets. As doping is increased, dc conductivity increases and again falls with further increase in doping. This fall is believed to be due to the decrease in the number of vacancies, reducing the number of charge carriers in the system \cite{26}. The result is very much qualitatively similar in nature to that of ours (see Fig.(3)), which shows that the charge stiffness also shows a peak around the optimal doping region and a sharp decrease as $\delta$  is increased further. \\
In the present sub-section, we have also tried to compare our results of charge stiffness with effective Coulomb interaction for doped systems. In this context, it must be pointed out that no direct experimental results are available for effective Coulomb interaction (V$_{eff}$) of layered cuprate systems. So, one can extract V$_{eff}$ from results of optical experiments, using the constitutive equations as given below. V$_{eff}$ in the long wavelength limit of the staggered magnetization is related to the imaginary conductivity by the standard constitutive equations in the continuum limit \cite{43}: 
\begin{align}
\epsilon^{\prime}(\omega)=1-\frac{4\pi\sigma^{\prime\prime}}{\omega}
\end{align}
Thus, the effective Coulomb interaction.
\begin{align}
V_{eff}(\omega)=\frac{V_0}{\epsilon^{\prime}(\omega)} 
\end{align}
leading to
\begin{align}
V_{eff}(\omega)=\frac{V_0}{1-\frac{4\pi \sigma^{\prime \prime}}{\omega}}
\end{align}
and using eq.(28) in the very low frequency limit,
\begin{align}
V_{eff}(\omega)=\frac{V_0}{1-\frac{4\pi D_c}{\omega^2}} 
\end{align}
with V$_0$ being the bare Coulomb interaction.\\

$\epsilon^{\prime}$ is the real part of the dynamic dielectric function and $\sigma^{\prime\prime}$ represents the imaginary part of the dynamic conductivity, which can be extracted experimentally. \\
\hspace*{0.3cm} The most of the experiments carried out on the planes of lightly and optimally doped La$_{2-x}$Sr$_x$CuO$_4$ are at high frequency and at much higher temperatures ($>>$0K), which are not suitable for comparison with our results. However, here, we have considered a transmitted THz time-domain spectroscopy (THz-TDS) on La$_{2-x}$Sr$_x$CuO$_4$\cite{32}. The effective Coulomb interaction is derived from the experimentally extracted imaginary conductivity using eq.(32). We have found that the effective Coulomb interaction is small and remains almost constant throughout the lower doping region (in the calculation, we have used the bare onsite Coulomb interaction V$_0$=3.5eV in the undoped phase \cite{32}). This result is similar to that of our derived charge stiffness constant as a function of doping in the under-doped region. Moreover, the eq.(33) shows the possibility of V$_{eff}(\omega)$ turning attractive for $\omega\rightarrow$0; assuming D$_c\propto$D in the RPA-like treatment of correlated phase, even with the values of D$_c$, as allowed by stability criterion. Thereafter, we are awaiting our theoretical prediction of effective Coulomb repulsion to be directly tested by experiments in near future.

\section{Discussion}
The generalized charge stiffness constants for 2D and 1D t-J-like models in strong and weak correlation limits are calculated. A weak dimensional dependence is seen for coupling between the mobile charge degrees of freedom. Furthermore, our calculations bring out several important features and conclusions covering various aspects of correlated fermionic systems in low dimensions. These are discussed below in detail: 
\subsection{Equivalence of generalized charge stiffness constant with Drude weight and effective Coulomb interaction}
\hspace*{0.3cm} The D$_c$ in 2D remains zero upto $\delta$=$\delta_c$=0.61 and then exhibits a sharp rise in value. $\delta_c$ shifts to optimal doping region when t$_2$ and t$_3$ are included. The effective Drude weight (D) also shows a similar kind of behaviour as shown in the previous section (see Fig.(5)) \cite{26}. In the low doping region, the Drude weight remains zero, signifying the insulating behaviour of the antiferromagnets. Further, the rise in the magnitude of `D' with doping indicates the rise in the number of mobile holes, however, if the doping is still increased, one can observe a fall in `D'. As we have already mentioned that, this fall is due to the inclusion of large number of vacancies in the system, which greatly suppresses the number of charge carriers, thus resulting in the fall of Drude weight \cite{26}.\\
Moreover, in this paper, we have also tried to establish a connection between our derived D$_c$ and effective Coulomb interaction in doped antiferromagnetic systems, within the RPA. The V$_{eff}$ extracted from experimental data shows that the effective Coulomb interaction remains almost constant in the lower doping region, which is very similar to the behaviour of our derived charge stiffness in the entire under-doped regime (see Fig.(3)) \cite{32}. The characteristic behaviour of the coupling between the charge degrees of freedom in the low doping regime is quite physical. In the under-doped regime, the correlation is very strong with $\alpha$=1, preventing the two carriers from approaching close to each other and thus largely suppressing the itinerant behaviour of the charges. As a result, the Drude weight is very small and the charges are far apart to feel the mutual repulsion, giving a zero value to effective Coulomb interaction, which remains constant throughout the lower doping region. When the doping concentration is gradually increased, the charge degrees of freedom become mobile and they can feel the repulsive interaction due to the other charges, as long as the $\delta$ do not become high so as to screen the Coulomb repulsion between them.
\subsection{Effective Coulomb interaction for high density electron gas}
In the medium and the over-doped regime, where the correlation weakens and the charges become mobile, one can take the continuum approximation and observe a point of discontinuity in the Lindhard function at q=2k$_F$ (`q' is the charge ordering wave vector). Then, using different values of the ordering wave vector `q', it can be shown that the discontinuity appears at some value of $\delta$ in the optimal doping region (Fermi momentum being related to $\delta$ by eq.(16)) \cite{44,45}. This discontinuity in the Lindard function also manifests itself in the calculation of dielectric function and as a result V$_{eff}$ shows a jump at the corresponding value of doping concentration \cite{44,45} (see Appendix). This characterisctic behaviour is very similar to our result of derived charge stiffness constant (see Fig.1), which possibly signifies the tendency of the formation of charge ordering or charge density waves as the idea put forward by Overhauser \cite{46}. Hence, the similarity in the behaviour D$_c$ constant and V$_{eff}$ proves the qualitative equivalence between the two, even in the over-doped regime of these doped itinerant systems. Considering the equivalence, we have drawn a phase diagram of the doped antiferromagnets in 2D, based on their charge responses from the t$_1$-t$_2$-t$_3$-J model (see Fig.(6)). We have shown the values of critical doping concentration ($\delta_c$) for different values of t$_3$/t$_1$ ratio, taking t$_2$/t$_1$ as parameter. One can also notice that for a particular value of t$_3$/t$_1$, the transition between the two regions of different charge couplings, takes place at a lower value of doping concentration for higher values of $\mid$t$_2$/t$_1\mid$.
\begin{figure}[H]
\centering
\includegraphics[scale=0.85]{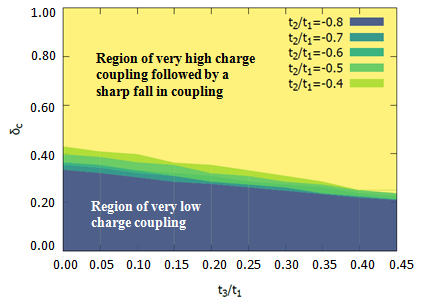}
\caption {Phase diagram showing the critical doping concentration ($\delta_c$), separating the regions of charge couplings, as a function of t$_3$/t$_1$ (with t$_2$/t$_1$ ratio as the parameter). The regions of doping concentration below $\delta_c$ represent the regime very low charge coupling and above $\delta_c$, the interaction shows a very high value, followed by a sharp fall. The different colours are used for different ratios of t$_2$/t$_1$ [$\alpha$=1 has been taken]. }
\end{figure}
\subsection{Comparison between behaviours of D$_c$ and D$_s$ in 2D} The exchange energy contribution to D$_c$ vanishes in the entire doping region (see eq.(12)), resulting in the distinct behaviour of spin and charge stiffness constants \cite{20}. We have shown that D$_c$ for two-dimensional lattice with pure t-J model remains zero throughout the lower doping region and exhibits a sharp rise at $\delta$=$\delta_c$=0.61. After this point, D$_c$ immediately falls as doping is increased further (see Fig.(1)). The parameter $\delta_c$ shifts to the optimal doping region when two higher neighbour hoppings are included (see Fig.(3)). \\
\hspace*{0.3cm} The region of enhanced Coulomb interaction around $\delta_c$ may imply a tendency towards the formation of a charge density wave, as described in the previous sub-section \cite{45,46}. Some of the previous theoretical results also confirms the presence of charge density wave states in the context of single-band t-J-U model \cite{47}. Interestingly, the spin stiffness constant (D$_s$) also shows a point of inflection (indicative of a possible phase transition) at the same $\delta$, where D$_c$ exhibits the sharp rise \cite{20}. Furthermore, D$_c$ and D$_s$ show almost identical behaviour for $\delta>\delta_c$, i.e. in the over-doped regime, which is an expected behaviour of Fermi liquid-like phases. In the under-doped regime however, the behaviour of the two stiffness constants are very different. Thus it can be concluded that we get the two regions of distinctly different behaviours. The regions are very likely to characterize (i) an anomalous conducting phase and (ii) a Fermi liquid-like metallic phase. 
\subsection{Comparison between behaviours of D$_c$ and D$_s$ in 1D} 
For t-J model, the quantity D$_c$ in 1D vanishes at $\delta$=0 and $\delta\rightarrow$1 and exhibits a maximum in the lower $\delta$ region. The peak shifts to further lower doping as the higher neighbour hopping amplitudes are increased and reaches the $\delta\rightarrow$0 limit at critical values of t$_2$ and t$_3$(see Figs.(2),(4)). In a recent paper, we have shown that in one dimension, D$_s$ displays a high value at $\delta\rightarrow$0 limit and falls rapidly with increase in doping concentration \cite{21}. The drastic fall is immediately followed by the formation of a peak in the under-doped regime \cite{21}. Hence, we see that D$_s$ and D$_c$ show completely distinct behaviour only in the very low doping region, whereas they show a similar trend as doping is slightly increased. \\
\hspace*{0.3cm} Furthermore, it is also seen that the tendency towards the formation of charge density wave occurs at much lower doping concentration in 1D than in 2D. Thus the dimensional dependence of charge stiffness in low dimensional systems is also established, similar to the spin stiffness case.  
\subsection{Possibility of pair formation} 
In some of the recent works, real space pairing has been studied in the framework of the t-J-like models mostly in the under-doped phase \cite{48,49,50,51,52}. In our calculation we do not get any region of negative charge stiffness, as is expected from the stability criteria (see eqs.(2)-(5)). However, eq.(33) shows that the effective Coulomb interaction can be attractive in a range of doping concentration, where charge stiffness constant has a large non-zero value. This signifies the possibility of superconducting pair formation in a region where D$_c$ shows a peak, i.e., the region of optimal opting concentration (see Fig(3)). \\
\newpage
\appendix
%\numberwithin{equation}{section}
\section*{Appendix}
\section*{Calculation of effective interaction from dielectric function approach}
 The longitudinal electronic dynamic dielectric function for a weakly correlated Fermi liquid-like phase for band electrons can be expressed as\cite{22,44,45}:
 \setcounter{equation}{0}
 \renewcommand{\theequation}{A.\arabic{equation}}
\begin{align}
\epsilon^{-1}(\overline{q}+\overline{G},\overline{q}+\overline{G^\prime},\omega)=1+V_{0}(\overline{q}+\overline{G}) \chi(\overline{q}+\overline{G},\overline{q}+\overline{G^\prime},\omega)
\end{align}
where $\overline{G}$ and $\overline{G^\prime}$ are Umklapp vectors corresponding to the lattice background  and in 2D \cite{44}\\
\begin{align}
   V_0(\overline{q}+\overline{G})=\frac{2\pi e^2}{\vert\overline{q}+\overline{G}\vert}
   \end{align}
   is the bare Coulomb interaction between the electrons, projected in a 2D layer.\\
At the conventional RPA level $\chi$, the screened dynamic charge susceptibility neglecting the exchange-correlation effects, is given by \cite{44,45}:
\begin{align}
\chi(\overline{q}+\overline{G}, \overline{q}+\overline{G^\prime},\omega)=\frac{\chi_0(\overline{q}+\overline{G}, \overline{q}+\overline{G^\prime},\omega)}{1-V_0\chi_0(\overline{q}+\overline{G}, \overline{q}+\overline{G^\prime},\omega)}	
\end{align}
where, $\chi_0$($\overline{q}$+$\overline{G}$, $\overline{q}$+$\overline{G^\prime}$,$\omega$) is the free charge dynamic susceptibility given by the Lindhard function \cite{45}. \\
Hence, the effective static Coulomb interaction obeys the equation:
\begin{align}
\frac{1}{V_{eff}(\overline{q}+\overline{G^\prime},0)}=-\chi_0(\overline{q}+\overline{G}, \overline{q}+\overline{G^\prime},0)+\frac{1}{V_0(\overline{q}+\overline{G})}
\end{align}

%\begin{acknowledgements}
%\end{acknowledgements}




\end{document}